\documentclass[a4paper,11pt]{article}
\usepackage{jinstpub}

\usepackage{lineno}
\usepackage{siunitx}
\usepackage{glossaries}
\usepackage{subcaption}
\usepackage[version=4]{mhchem}
\usepackage{multirow}
\usepackage{makecell}

\DeclareSIUnit{\u}{u}
\DeclareSIUnit{\samples}{Sa}

\newacronym{HF}{HF}{high frequency}
\newacronym{SiC}{SiC}{silicon carbide}
\newacronym{RF}{RF}{radio frequency}
\newacronym{FWHM}{FWHM}{full width at half maximum}
\newacronym{EBC}{EBC}{empty bucket channeling}
\newacronym{HEP}{HEP}{high-energy physics}
\newacronym{PATD}{PATD}{particle arrival time distribution}
\newacronym{MPV}{MPV}{most probable value}
\newacronym{PCB}{PCB}{printed circuit board}

\title{Measurements of the micro-spill structure of medical cyclotron and synchrotron beams and its impact on pulse pileup}

\author[a,1]{Matthias Knopf, \note{Corresponding author.}}

\author[b]{Simon Waid,}
\author[b]{Stefan Gundacker,}
\author[b,a]{Sebastian Onder,}
\author[b,a]{Daniel Radmanovac,}
\author[b]{Philipp Gaggl,}
\author[c,d]{Giulio Bordieri,}
\author[c,e]{Francesco Cordoni,}
\author[c,f]{Marta Missiaggia,}
\author[c]{Enrico Verroi,}
\author[g]{Giulio Magrin,}
\author[b]{Thomas Bergauer,}
\author[a]{Albert Hirtl}

\affiliation[a]{TU Wien, Atominstitut,\\Stadionallee 2, 1020 Wien, Austria}
\affiliation[b]{Marietta Blau Institute for Particle Physics (MBI), Austrian Academy of Sciences,\\Dominikanerbastei 16, 1010 Vienna, Austria}
\affiliation[c]{Trento Institute for Fundamental Physics and Applications (TIFPA), INFN,\\Via Sommarive, 14, 38123 Povo, TN, Italy}
\affiliation[d]{University of Trento, Department of Physics,\\Via Sommarive, 15, 38123 Povo, TN, Italy}
\affiliation[e]{University of Trento, Department of Civil, Environmental and Mechanical Engineering,\\Via Mesiano, 7, 38123 Povo, TN, Italy}
\affiliation[f]{Louisiana State University, Department of Physics,\\202 Nicholson Hall, Baton Rouge, LA 70803}
\affiliation[g]{EBG MedAustron GmbH,\\Marie-Curie Straße 5, 2700 Wiener Neustadt, Austria}

\emailAdd{matthias.knopf@tuwien.ac.at}

\abstract{Detector characterization and instrumentation testing are often performed at cyclotron and synchrotron facilities, many of which were originally developed for medical applications in cancer therapy. For particle physics experiments requiring a single-particle resolution, pileup can significantly degrade data quality, making precise knowledge of the beam time structure essential for selecting appropriate readout parameters. However, such information is often unavailable from the facilities and challenging to determine experimentally. Here, we report measurements of the spill time structure at two medical accelerator facilities using a silicon carbide (SiC) particle sensor coupled to a high-frequency readout system. Owing to its high carrier saturation velocity and the tolerance to large bias voltages, SiC is well suited for fast readout and measurements requiring precise timing. Using \color{black} an analog readout circuit with \SI{6}{\giga\hertz} bandwidth and \color{black} custom SiC diodes, we characterize the micro-spill structure of both cyclotron and synchrotron beams on a sub-nanosecond timescale. The measured arrival-time distributions exhibit modulation with the accelerator RF frequencies, reflecting features of the extraction process. The resolved micro-spill structure enables quantitative estimation of pileup contributions and provides design constraints for future readout electronics. The presented results emphasize the importance of the characterization of the beam time-structure characterization for the development of precise readout systems.}


\arxivnumber{2605.07508}

\begin{document}
\maketitle
\flushbottom

\section{Introduction}
\label{sec:intro}

Cancer therapy using light ions has become a valuable and steadily growing alternative to conventional photon therapy, supported by a global network of cyclotron- and synchrotron-based treatment facilities currently in clinical operation~\cite{PTCOG}. Several of the accelerator facilities originally designed for cancer therapy are open to external researchers and are routinely used for research, e.g. to test instrumentation and radiation detectors for \gls{HEP}, medical physics and space research~\cite{MedAustron,Trento,CNAO}. Some of these measurements are sensitive to the time structure of the delivered beams in terms of pileup. This information is often unknown and not obtainable from the existing monitoring systems at the facilities, particularly at small timescales not relevant for therapeutic applications (>\SI{}{\kilo\hertz}). During treatment, the delivered dose is most commonly monitored using parallel plate ionization chambers in the nozzle~\cite{Giordanengo_2013}. Typical systems are limited in both charge resolution ($\sim$\SI{100}{\femto\coulomb}) as well as time resolution ($\sim$\SI{100}{\micro\second})~\cite{Giordanengo_2015} and are not capable of characterizing the particle beam on a single-particle basis~\cite{Ulrich-Pur_2021}. The particle ﬂux of accelerator beams reﬂects the time structure of the radio frequency ﬁeld, which is typically on the order of \SIrange{1}{100}{\mega\hertz}.\\

Solid-state detectors, typically manufactured from silicon, offer the required sensitivity and timing capabilities to register individual particle crossings at clinical rates ($\sim$\SI{e9}{\per\second}~\cite{ICRU_93}) due to the comparatively low charge collection times <\SI{1}{\nano\second}. In recent years, \gls{SiC} has emerged as a promising sensor material for fast, time-resolved measurements due to its unique combination of properties as a wide bandgap material. Its high breakdown field and high carrier saturation velocity enable fast readout capabilities. When paired with multi-GHz electronics, pulse duration well below \SI{1}{\nano\second} can be achieved. These characteristics make \gls{SiC} particularly attractive for beam monitoring in ion therapy, where high temporal resolution is essential.\\

This paper presents an analysis of the micro-spill structure of a medical cyclotron and synchrotron facility, obtained with a custom \gls{HF} \gls{SiC} readout \cite{Gsponer_2025}, complementing previous measurements~\cite{Knopf_2025}. The arrival times between successive particles, the \glspl{PATD}, are compared and the implications for pileup-sensitive measurements are discussed.\\

\section{Materials and Methods}

\paragraph{Pileup} 

Accurate tracking in counting measurements, such as time-of-flight detectors or particle counters, requires pileup-free signal acquisition. Similarly, in spectroscopic measurements, precise energy spectra depend on the reliable detection and analysis of pulse amplitudes. However, at high particle rates, types of measurement suffer from reduced accuracy due to the overlap of individual events in the analog signal. A readout system is defined by a characteristic processing time, $\tau$, typically set by the shaping or peaking time, during which events can be resolved independently. For high-resolution spectroscopic systems, this is typically on the order of \SI{1}{\micro\second}. Pulse pileup occurs when subsequent particles arrive within intervals $\Delta t <\tau$. This can result in the complete merging of two or more pulses, known as peak pileup, where the system registers only a single pulse. If the events remain partially separable, referred to as tail pileup, the measured pulse heights no longer accurately represent the individual events in spectroscopic measurements. For Poisson-distributed inter-arrival times $\Delta t$, as in radioactive decay, the pileup probability $p_\text{PU}$ is determined by the effective mean rate $\alpha_0$, being the average rate scaled to the portion of the source actually reaching the detector, and the processing time $\tau$ as

\begin{equation}
    p_\text{PU}=1-\exp{(\alpha_0\tau)}.
\end{equation}

In accelerator beams, this relation no longer holds, as the extracted particle current is influenced by various subsystems such as the \gls{RF} system and the magnets that maintain particle trajectories, which themselves are subject to stochastic fluctuations from the electronics.

\paragraph{Trento cyclotron}

The Trento Proton Therapy Center (Italy) is equipped with an IBA Proteus 235 system comprising a C230 cyclotron, a beam transport line with a passive energy selection system, and two clinical beamlines featuring rotating gantries in two separate rooms. In addition, two dedicated beamlines are available in one separate room for experimental research. The isochronous cyclotron accelerates protons up to a fixed energy of \SI{230}{\mega\electronvolt}. The accelerator is operated at a fixed \gls{RF} of $\sim$\SI{106}{\mega\hertz} on the fourth harmonic of the orbital frequency \cite{Review_Cyclotrons}. The energy delivered to the treatment rooms is selected between nominally \SI{70.2}{\mega\electronvolt} and \SI{228.2}{\mega\electronvolt} using rotating passive degraders. The beam current at extraction (before energy selection) can be varied between \SIrange{1}{320}{\nano\ampere}. This corresponds to particle fluxes between \SIrange{e6}{e11}{\per\second} on Gaussian beamspots between \SIrange{2.5}{7}{\milli\meter} \gls{FWHM}~\cite{Trento}. In the experimental room, the delivered current can be additionally modulated to extract custom duty cycles. The beam passes through a \SI{70}{\micro\meter} titanium window before reaching the iso-center, located \SI{1.25}{\meter} downstream of the exit window, resulting in a small energy loss from the nominal value \cite{Trento}.\\

\paragraph{MedAustron synchrotron} 
MedAustron is an ion therapy and research facility located near Vienna (Austria). The center operates three beamlines for clinical use and an additional beamline dedicated to research. The center is based around a medical synchrotron derived from the Proton-Ion Medical Machine Study (PIMMS) design~\cite{PIMMS}. This synchrotron has a circumference of $L$=\SI{77.65}{\meter} and enables the acceleration of protons in the energy range from \SIrange{62.4}{800}{\MeV}, \ce{^12C^6+} ions from \SIrange{120}{402.8}{\MeV\per\u}~\cite{Pivi_2019_Status_Carbon_Commissioning}, and \ce{^4He^{2+}} ions from \SIrange{39.8}{402.8}{\MeV\per\u}~\cite{Gambino_2024_Helium}. In medical mode, particles are extracted from the synchrotron ring using betatron core–driven third-order resonant extraction ~\cite{Pullia_2016_Betatron_Extraction}, resulting in quasi-continuous spills between \SIrange[]{1}{10}{\second}. Typical particle fluxes delivered to the irradiation rooms range from \SIrange{e9}{e10}{\per\second} for protons and from \SIrange{e8}{e9}{\per\second} for carbon ions. To suppress intensity fluctuations (ripples) in the extracted beams from the power converters of the magnets, MedAustron uses \gls{EBC} in clinical mode~\cite{Kuehteubl_Diss,Crescenti_1998_EBC}. This introduces a strong modulation of the extracted beams with the synchrotron \gls{RF} frequency $f_\text{RF}\approx\beta c/L$, which corresponds to \SIrange{1}{3}{\mega\hertz} depending on the energy and particle species \cite{Knopf_2025}.

\paragraph{SiC-based \gls{HF} readout} 

\Gls{SiC} is a wide bandgap material (\SI{3.26}{\electronvolt}) gaining interest in the \gls{HEP} community. Due to a combination of high breakdown voltage and high carrier saturation velocity~\cite{DeNapoli_2022}, it enables fast timing performance. However, measuring fast transients with a solid-state detector requires not only a high bias voltage and an appropriate detector material, but also depends on detector geometry, a suitable amplifier, and proper interconnects. The combination of detector capacitance, amplifier input impedance and the inductance of the input line act as an RLC low-pass filter. The detector capacitance and the length of the bond wires and \gls{PCB} traces should therefore be minimized. To reduce pileup at high rates, the sensor should be coupled to a fast multi-\SI{}{\giga\hertz} amplifier and designed with a small surface area or appropriate segmentation.\\

\begin{figure}[htbp]
\centering
\includegraphics[width=.4\textwidth]{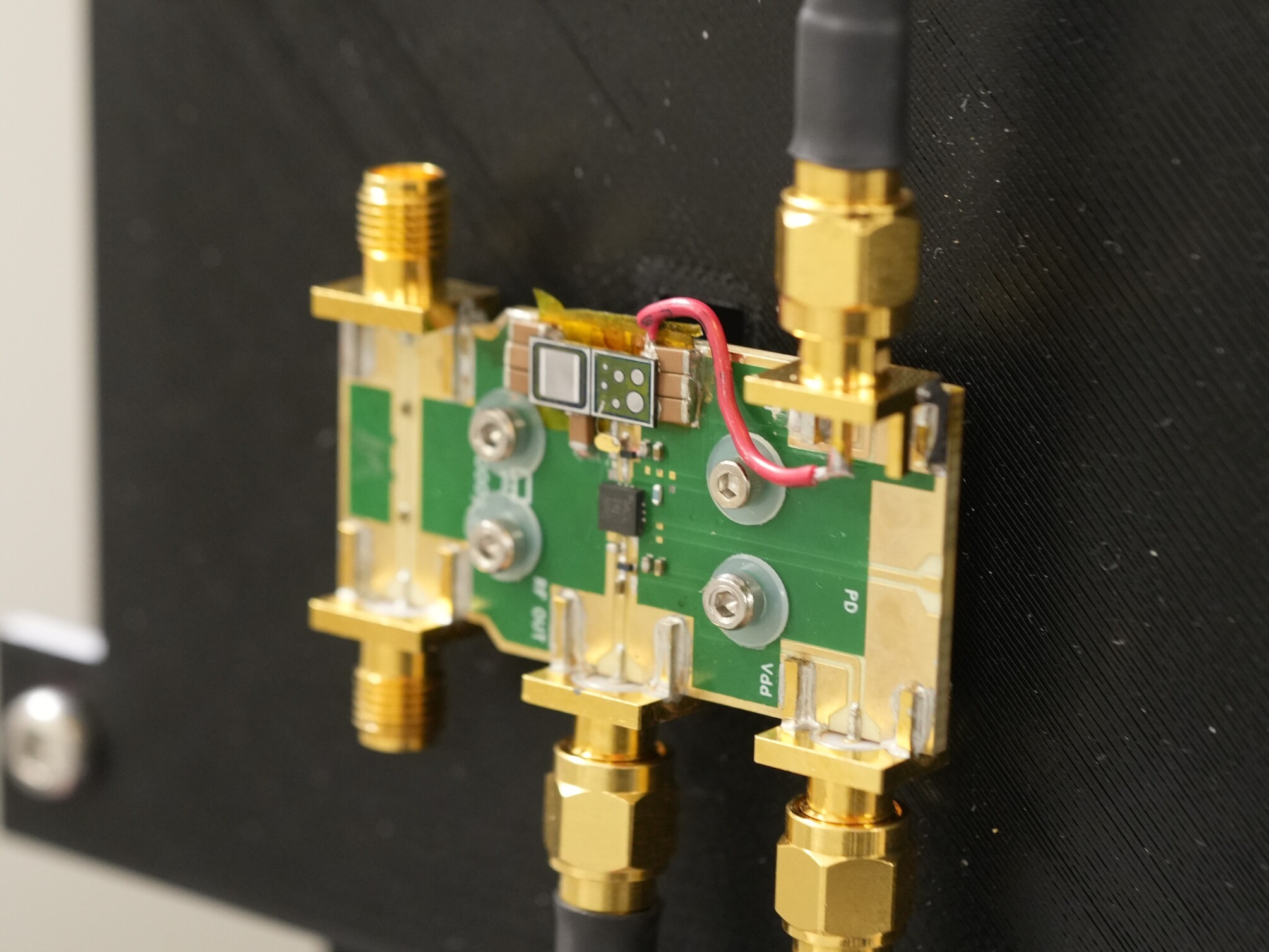}
\qquad
\includegraphics[width=.45\textwidth]{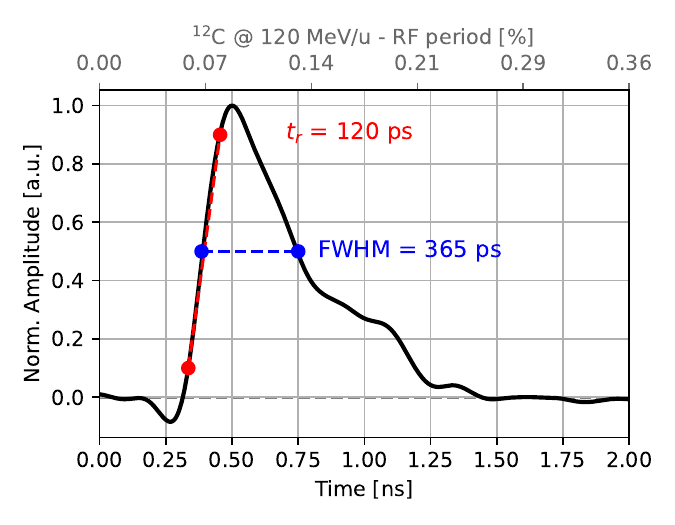}
\caption{\Gls{HF} readout board with the mounted \gls{SiC} sensor and an averaged waveform for \ce{^12C^6+} ions. \color{black}The transient shows the signature of a single-particle crossing obtained from the average of normalized waveforms. A secondary x-axis is included to provide a comparison of the signal duration with the time scale defined by the \SI{120}{\mega\electronvolt\per\u} \gls{RF} frequency.\color{black}\label{fig:HF_Board}}
\end{figure}

A small circular 4H-SiC p-i-n diode with \SI{140}{\micro\meter} diameter and \SI{50}{\micro\meter} active thickness was used for all measurements. A capacitance of \SI{70}{\femto\farad} has been measured at \SI{1}{\kilo\volt} bias. The sensor was designed by the authors and manufactured by IMB-CNM-CSIC in Barcelona (Spain). The readout is built around a Mini-Circuits PMA3-14LN+ low-noise amplifier exhibiting a flat gain of \SI{22.6}{\decibel} between \SI{50}{\mega\hertz} -- \SI{10}{\giga\hertz}. An additional Mini-Circuits ZX60-14LN-S+ low-noise amplifier with a gain of \SI{22}{\decibel} was used to boost the signal for digitization using a Rhode \& Schwarz RTO6 oscilloscope. \color{black} The effective bandwidth of the presented measurements is determined by the \SI{4}{\giga\hertz} analog bandwidth of the oscilloscope input, while the analog front-end electronics could provide a bandwidth of \SI{6}{\giga\hertz}. \color{black} The bond wires were kept as short as possible while ensuring adequate spacing to prevent HV breakdown on the \gls{PCB}. The digitized signal has a risetime of \SI{120}{\pico\second} and a \gls{FWHM} of \SI{365}{\pico\second}. The waveform reflects signal formation in the sensor. The initial slope arises from mirror charges building up on the readout electrode as ionization frees charge carriers in the sensitive volume. This is followed by electron drift up to \SI{400}{\pico\second} after the peak maximum, producing the first fall-off, and hole drift, which creates the second, flatter decline in the transient. Figure \ref{fig:HF_Board} shows the setup in the iso-center and an averaged waveform for \SI{402.8}{\mega\electronvolt\per\u} \ce{^12C^6+} -ions, acquired with the RTO6 oscilloscope at \SI{20}{\giga\samples\per\second}. Details about the readout can be found in \cite{Gsponer_2025}.

\paragraph{Measurements}

The data were acquired in three separate measurement campaigns using the same sensor and readout. Two measurement campaigns at the MedAustron synchrotron, with and without \gls{EBC}, and a third dataset at the Trento cyclotron. Each time, the sensor was placed in the iso-center and a bias voltage of \SI{1}{\kilo\volt} was applied using a Keithley 2470 SMU. Due to the limited oscilloscope memory (\SI{2}{\giga\samples} per shot), digitization was performed in \SI{200}{\milli\second} frames. At both MedAustron campaigns, the window was shifted at each extraction trigger signal with overlapping intervals to ensure full coverage of each spill and enable an analysis over the duration of \SI{10}{\second} spills (see figure \ref{fig:MAUS_Intensity}). Since the cyclotron extraction was continuous, the \SI{200}{\milli\second} frames were self-triggered on the detector signal in the Trento dataset. At MedAustron, data were acquired for the highest and lowest clinically relevant proton energies (\SI{62.4}{\mega\electronvolt} and \SI{252.7}{\mega\electronvolt}) and carbon-ion energies (\SI{120}{\mega\electronvolt\per\u} and \SI{402.8}{\mega\electronvolt\per\u}) with \gls{EBC} on and for \SI{120}{\mega\electronvolt\per\u} and \SI{402.8}{\mega\electronvolt\per\u} \ce{^12C^6+} ions as well as \SI{62.4}{\mega\electronvolt} protons without \gls{EBC}. The measurements with \gls{EBC} on were acquired at \SI{3.33}{\giga\samples\per\second}, while those without \gls{EBC} were acquired at \SI{10}{\giga\samples\per\second}. At the Trento cyclotron, the particle beam was extracted in a pulsed mode with \SI{10}{\second} intervals and a \SI{500}{\milli\second} duty cycle, to minimize detector activation during the dead time while saving data.  Data for the cyclotron proton beam were acquired at \SI{10}{\giga\samples\per\second} in four different machine settings at nominally \SI{70.2}{\mega\electronvolt} and beam currents of \SI{30}{\nano\ampere}, \SI{150}{\nano\ampere} and \SI{300}{\nano\ampere}, as well as nominally \SI{148.5}{\mega\electronvolt} protons at \SI{30}{\nano\ampere}.

\paragraph{Analysis}

Timestamps for the individual particle crossings are extracted from the digitized pulsetrain by identifying the position of the amplitude maximum associated with each threshold crossing in the signal within a \SI{1}{\nano\second} window around the initial crossing point. The threshold was set at $5\cdot \sigma$ of the baseline. In the Trento dataset, a \SI{50}{\mega\hertz} high pass filter was required to filter baseline fluctuations due to \gls{RF} pickup in the experimental room below the frequency band of the amplifier (\SI{50}{\mega\hertz} - \SI{10}{\giga\hertz}). With a rise time of \SI{120}{\pico\second}, at \SI{3.33}{\giga\samples\per\second}, a registered threshold crossing falls randomly within one of two possible time samples, yielding an uncertainty of $(2\cdot 300 \text{ ps})/ \sqrt{12}\approx$\SI{173}{\pico\second}, while at \SI{10}{\giga\samples\per\second}, the registered peak falls randomly into one of three possible time samples, leading to a timing uncertainty of $(3\cdot 100 \text{ ps})/ \sqrt{12}\approx$\SI{87}{\pico\second}. This is well below the minimum required keepout of \SI{1}{\nano\second}, to exclude tail pileup pulses. Pulses followed by large negative amplitudes are excluded from the analysis, since events of this polarity can not be triggered from particle crossings in the sensor and are likely produced by a particle hitting the amplifier. The time difference between consecutive particles $\Delta t$ is computed and used to derive the \gls{PATD} by binning the data. Normalized to unit area, these histograms represent the probability density function for the arrival time $\Delta t$ of the next particle crossing the sensor. The cumulative sum of the \gls{PATD} corresponds to the pileup probability for the given setup as a function of the characteristic processing time $\tau$ of a potential readout. The expectation value $E[\Delta t]$ of the \gls{PATD} can be calculated as

\begin{equation}
    E[\Delta t] = \frac{\sum_i\Delta t_i \cdot \text{PATD}(\Delta t_i)}{\sum_i \text{PATD}(\Delta t_i)}.
\end{equation}

\section{Results}

\paragraph{PATD - Particle Arrival Time Distributions}
\label{sec:Results}

\begin{figure}[htbp]
\centering
\includegraphics[width=.47\textwidth]{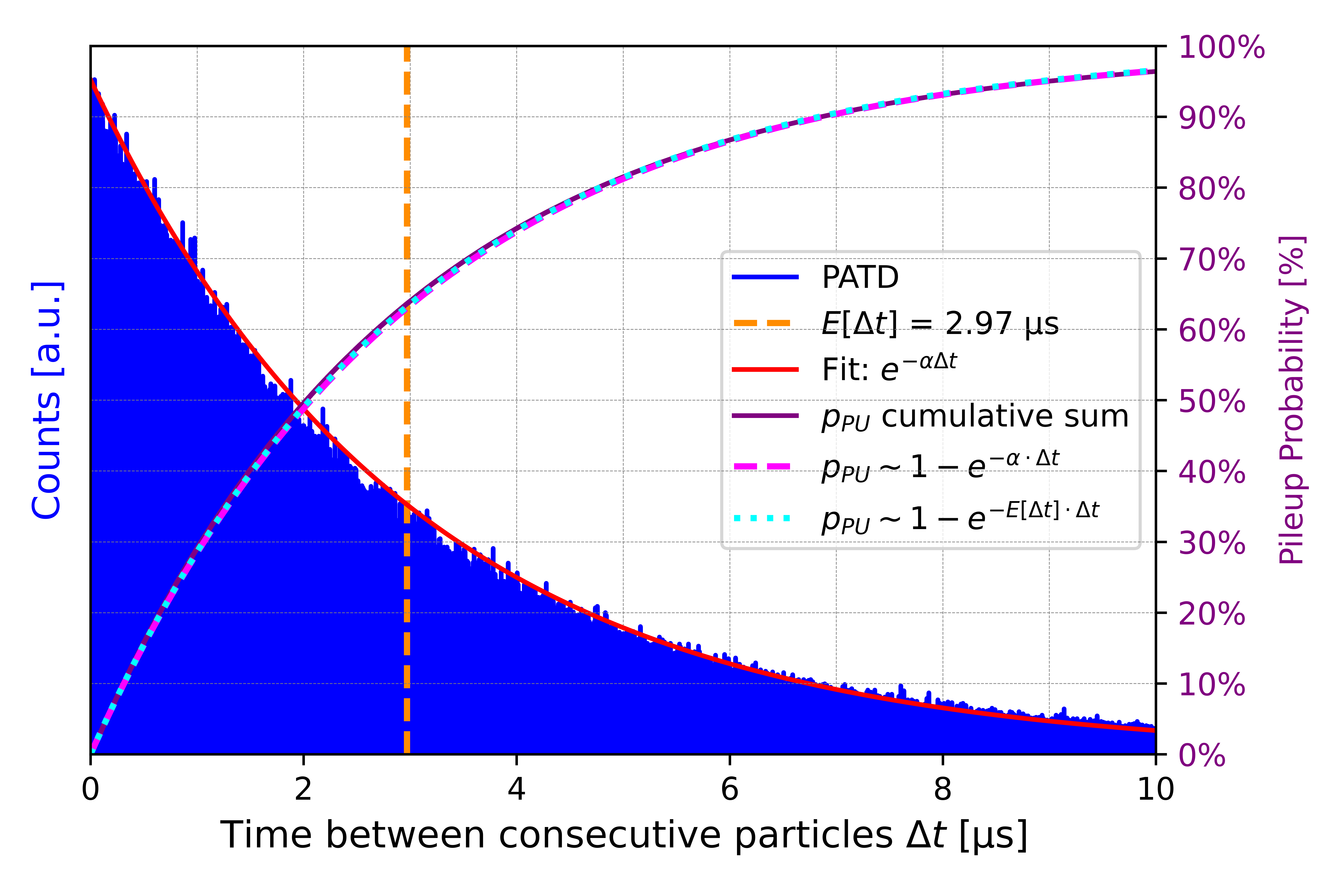}
\qquad
\includegraphics[width=.47\textwidth]{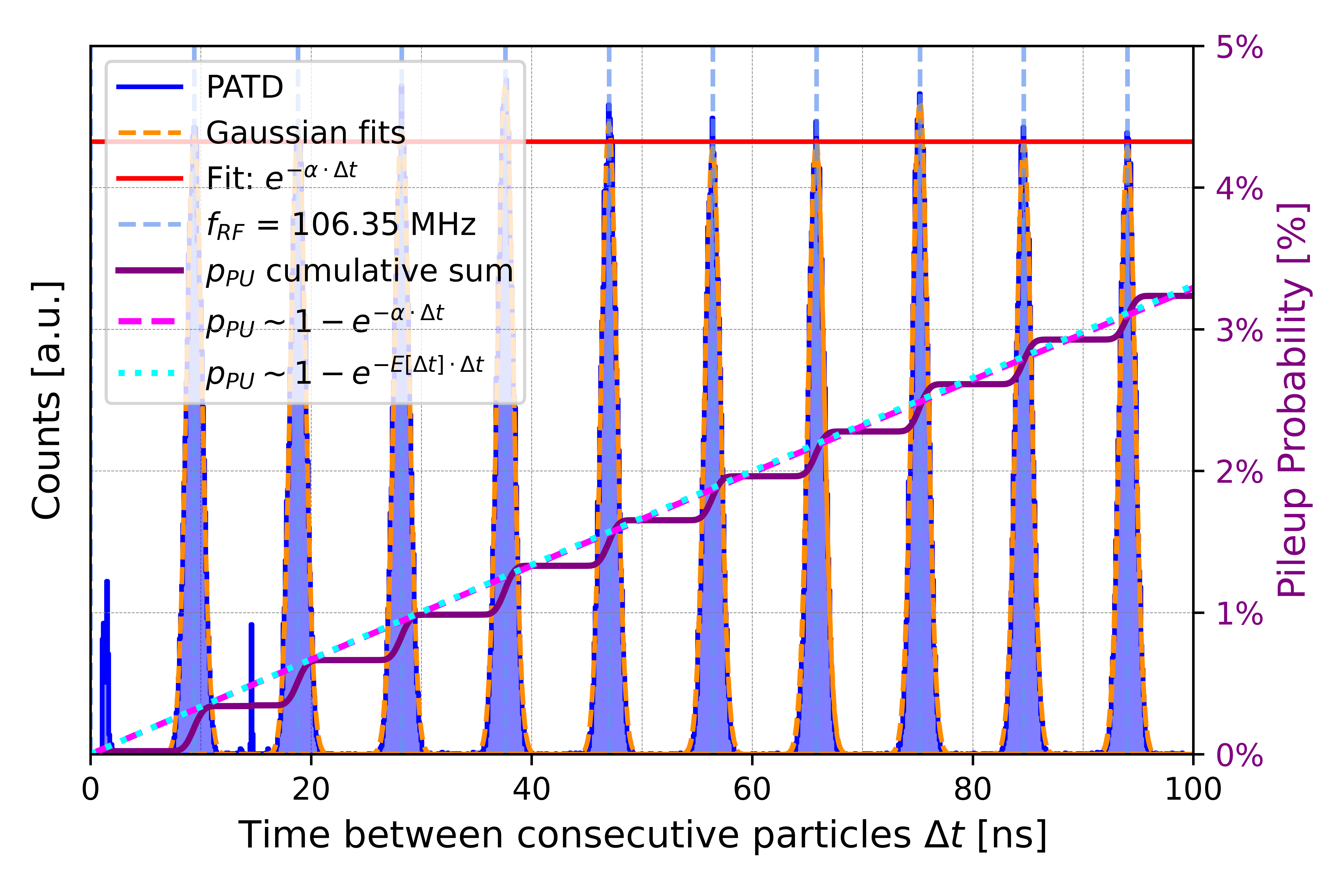}
\caption{\Gls{PATD} and pileup probability $p_{PU}$ for the \SI{148.5}{\mega\electronvolt} proton beam (\SI{30}{\nano\ampere}) at the Trento cyclotron. The distribution is modulated with the cyclotron \gls{RF} frequency $f_{RF}$=\SI{106.35}{\mega\hertz}. The plot on the right shows a zoomed section of the first 10 peaks.\label{fig:PATD_Trento_p148MeV}}
\end{figure}

Figure \ref{fig:PATD_Trento_p148MeV} shows the \gls{PATD} for the Trento cyclotron for \SI{148.5}{\mega\electronvolt} protons. Overall, the distribution shows an exponential decay, as expected from a Poisson distribution. However, the accelerator introduces a modulation with the \gls{RF} frequency ($f_\text{RF}=$\SI{106.35}{\mega\hertz}), resulting in a characteristic micro-spill structure, as illustrated in the right zoomed-in panel in figure \ref{fig:PATD_Trento_p148MeV}. As expected, all measurements at different particle currents and proton energies exhibit the same substructure, since the cyclotron initially accelerates all particles to \SI{230}{\mega\electronvolt} in a fixed \gls{RF} field. As to be seen in table \ref{tab:Trento_PATD}, the time constant of the exponential envelope decreases with increasing current and beam energy, since higher energies result in more focused beamspots.\\

% Details TRENTO
The individual peaks in the \glspl{PATD} were fit using Gaussian functions. The peak position and amplitude are determined respectively from the Gaussian mean $\mu_G$ and amplitude. These amplitudes are subsequently used to fit an exponential decay function $f(\Delta t)=\exp{(-\alpha\cdot \Delta t)}$. From the centers of the Gaussian fits $\mu_G$, a micro-bunch period of \SI{9.403}{\nano\second} was determined, corresponding to \SI{106.383}{\mega\hertz}. This is consistent with the reported cyclotron \gls{RF} frequency of \SI{106.35}{\mega\hertz}~\cite{Review_Cyclotrons}. The width of the micro-bunches were determined as $\sigma_G^{70.2 \text{ MeV}}=$\SI{1.32}{\nano\second} for the \SI{70}{\mega\electronvolt} beams and $\sigma_G^{148.5 \text{ MeV}}=$\SI{0.71}{\nano\second} for the \SI{148.5}{\mega\electronvolt} beam. Table \ref{tab:Trento_PATD} shows the results for the exponential fit values $\alpha$, the expectation values $E[\Delta t]$ and the average difference between the centers $\mu$ of the fitted Gaussians, as well as their average widths $\sigma_G$ for the Trento dataset. The indicated errors are the sample standard deviations over the first 7500 Gaussians per dataset (2000 for the \SI{30}{\nano\ampere} data). The large uncertainties in the Gaussian fit values of \SI{30}{\nano\ampere} dataset are attributed to limited counting statistics and these values were consequently excluded from the overall determination of the average micro-bunch spacing and width.\\

\begin{table}
\centering
\caption{Expectation values and fit parameters for the Trento cyclotron measurements.\label{tab:Trento_PATD}}
\begin{tabular}{l|cccc}
\textbf{Beam} &
\begin{tabular}{c}\textbf{Expectation value}\\\textbf{$E[\Delta t]$ [\SI{}{\micro\second}]}\end{tabular} &
\begin{tabular}{c}\textbf{Exponential fit }\\\textbf{$1/\alpha$ [\SI{}{\micro\second}]}\end{tabular} &
\begin{tabular}{c}\textbf{Diff. between}\\\textbf{$\mu_G$ [\SI{}{\nano\second}]}\end{tabular} &
\begin{tabular}{c}\textbf{Widths}\\\textbf{$\sigma_G$ [\SI{}{\nano\second}]}\end{tabular} \\
\hline
p, \SI{70.2}{\mega\electronvolt}, \SI{30}{\nano\ampere} & 49.89 & 42.59 & 9.383(374) & 2.29(1.93) \\
p, \SI{70.2}{\mega\electronvolt}, \SI{300}{\nano\ampere} & 4.87 & 4.66 & 9.403(60) & 1.32(4) \\
p, \SI{70.2}{\mega\electronvolt}, \SI{150}{\nano\ampere} & 11.03 & 10.76 & 9.403(93) & 1.31(6) \\
p, \SI{148.5}{\mega\electronvolt}, \SI{30}{\nano\ampere} & 2.97 & 2.89 & 9.403(30) & 0.71(2)
\end{tabular}
\end{table}

Due to the fast modulation, the expectation values $E[\Delta t]$ of the distributions are very similar to the decay time of their exponential envelope $1/\alpha$. The pile-up probabilities derived from $E[\Delta t]$, $\alpha$ and the cumulative sum of the \gls{PATD} are thus very similar, as shown in figure \ref{fig:PATD_Trento_p148MeV}. Since the modulation of the extracted beams occurs on a timescale well below the timing resolution of most applications, this suggests that the standard Poisson formulation of pileup statistics can be applied by interpreting the decay constant of the exponential envelope as the mean effective rate $\alpha_0$.\\

\begin{figure}[htbp]
\centering
\includegraphics[width=.47\textwidth]{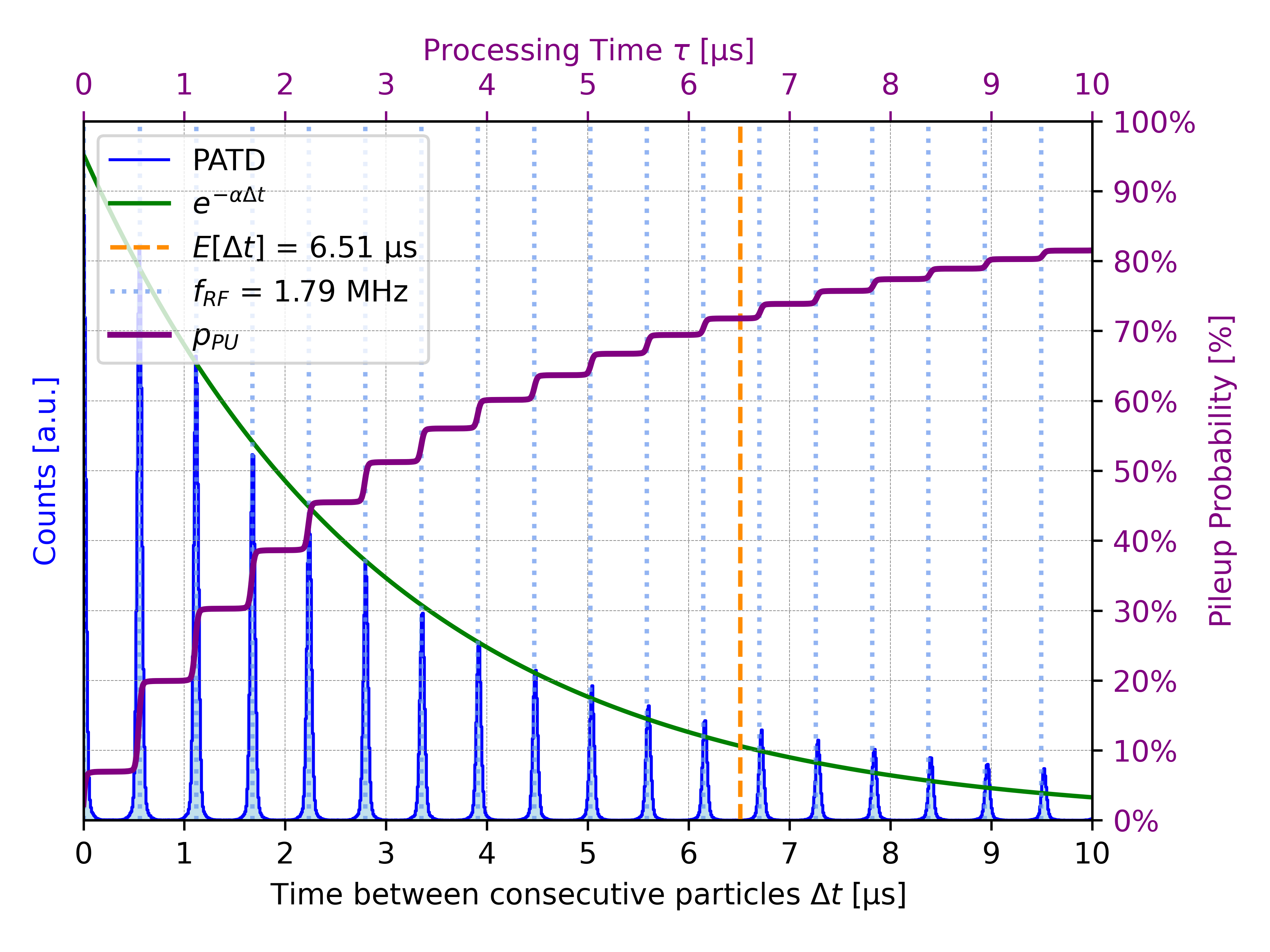}
\qquad
\includegraphics[width=.47\textwidth]{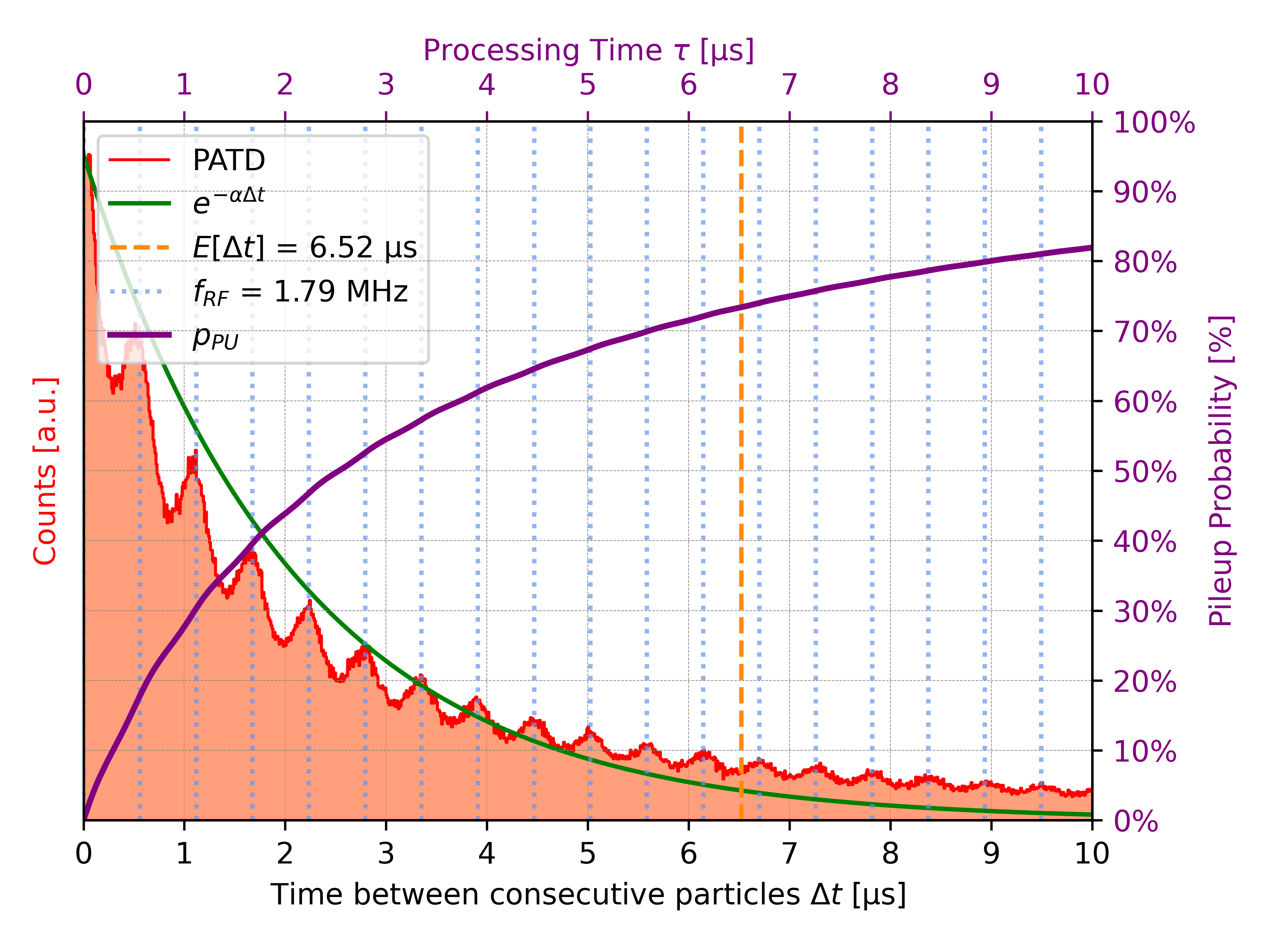}
\caption{\Gls{PATD} and pileup probability $p_{PU}$ for \SI{120}{\mega\electronvolt\per\u} \ce{^12C^6+} beams at the MedAustron synchrotron. The left plot shows the distribution with \gls{EBC} on, while the right plot shows a setting with \gls{EBC} turned off. The harmonics of the synchrotron \gls{RF} period are indicated in the plots.\label{fig:PATD_MedAustron_C120MeV}}
\end{figure}

% Allgemein MAUS
In a synchrotron, a more complex \gls{PATD} is observed, depending on the particle mass, energy and the mode of extraction, in particular on whether the \gls{RF} field is active during extraction or not. Figure \ref{fig:PATD_MedAustron_C120MeV} shows the \gls{PATD} for the MedAustron synchrotron for \SI{120}{\mega\electronvolt\per\u} \ce{^12C^6+} ions. The left panel shows the situation with \gls{EBC} on (\gls{RF} on during extraction) resulting in a strong modulation of the extracted current with the synchrotron frequency $f_\text{RF}=\beta c / L$. The right panel shows the \gls{PATD} for an extraction setting without \gls{EBC} (\gls{RF} off during extraction). The modulation is significantly weaker, although remnants are still observable as small peaks on top of the now filled-out distribution. In both cases, the envelope of the distributions is well described by an exponential at short inter-arrival times (up to $\Delta t \sim$\SI{5}{\micro\second}), but shows significant deviation from an exponential at larger time intervals. Since the relevant timescale for most detector systems is at this lower end, the distribution can be adequately described and scaled using an exponential fit of the envelope for the purpose of pileup estimation, retaining the high-frequency sub-structure.\\

No direct correlation between the pileup probabilities of the settings with and without \gls{EBC} can be established since the decay envelope of the distributions depends on the macroscopic beam current and other machine settings. The current is modulated with ripples in the \SI{}{\hertz} -- \SI{}{\kilo\hertz} range which can be interpreted as a change in the average beam current not affecting the micro-spill structure at the \SI{}{\micro\second} -- \SI{}{\nano\second} scale, but is becoming relevant at larger timescales, potentially explaining the observed deviation from an exponential envelope at time scales above $\sim$ \SI{5}{\micro\second}. \color{black} These ripples are a well-known characteristic of the MedAustron extracted beam and can still be observed in extracted beam measurements when \gls{EBC} is applied as a mitigation strategy~\cite{DeFranco_2018,DeFranco_2021}. \color{black} Due to the greater operational complexity of a synchrotron a direct comparison between the two beam configurations is not feasible as it is with a cyclotron. The measured distributions should therefore be viewed as references for a specific beam setting, which can be scaled to other readout electronics in that same beam. This procedure is described in section \ref{sec:Discussion}. Table \ref{tab:MAUS_PATD} shows the expectation values $E[\Delta t]$ for the MedAustron measurements with and without \gls{EBC}. Further analysis of the MedAustron data in the standard clinical mode with \gls{EBC} can be found in \cite{Knopf_2025}.\\

\begin{table}
\centering
\caption{Expectation values of the \glspl{PATD} for the MedAustron synchrotron beams.\label{tab:MAUS_PATD}}
\begin{tabular}{l|cc}
\textbf{Beam} &
\begin{tabular}{c}\textbf{Expectation value $E[\Delta t]$}\\\textbf{with \gls{EBC} [\SI{}{\micro\second}]}\end{tabular} &
\begin{tabular}{c}\textbf{Expectation value $E[\Delta t]$}\\\textbf{without \gls{EBC} [\SI{}{\micro\second}]}\end{tabular} \\
\hline
p, \SI{62.4}{\mega\electronvolt} & 12.01 & 22.57 \\
\ce{^12C^6+}, \SI{120}{\mega\electronvolt\per\u} & 6.51 & 6.52 \\
\ce{^12C^6+}, \SI{402.8}{\mega\electronvolt\per\u} & 5.38 & 4.16 \\
\end{tabular}
\end{table}

\begin{figure}[htbp]
\centering
\includegraphics[width=\textwidth]{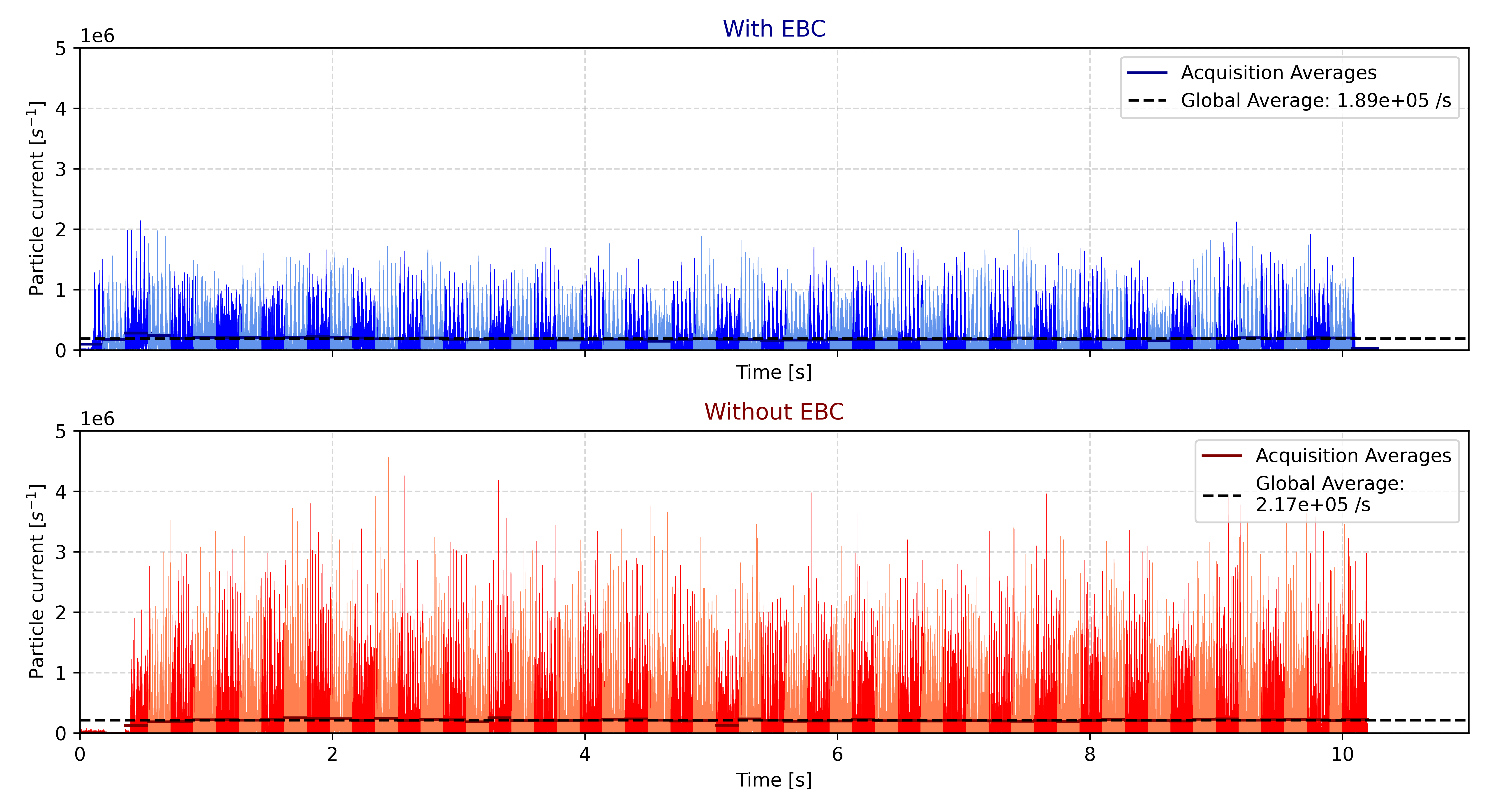}
\caption{Measured particle currents in the detector for the MedAustron \ce{^12C^6+} beam at \SI{402.8}{\mega\electronvolt\per\u} with \gls{EBC} (blue) and without \gls{EBC} (red). The current has been integrated into \SI{50}{\micro\second} bins. Different parts of the \SI{10}{\second} spills were acquired in \SI{200}{\milli\second} frames. Both the average rate and the fluctuations in the current are higher in the setting without \gls{EBC}. \label{fig:MAUS_Intensity}}
\end{figure}

\paragraph{Particle Current}

\begin{figure}[htbp]
\centering
\includegraphics[width=\textwidth]{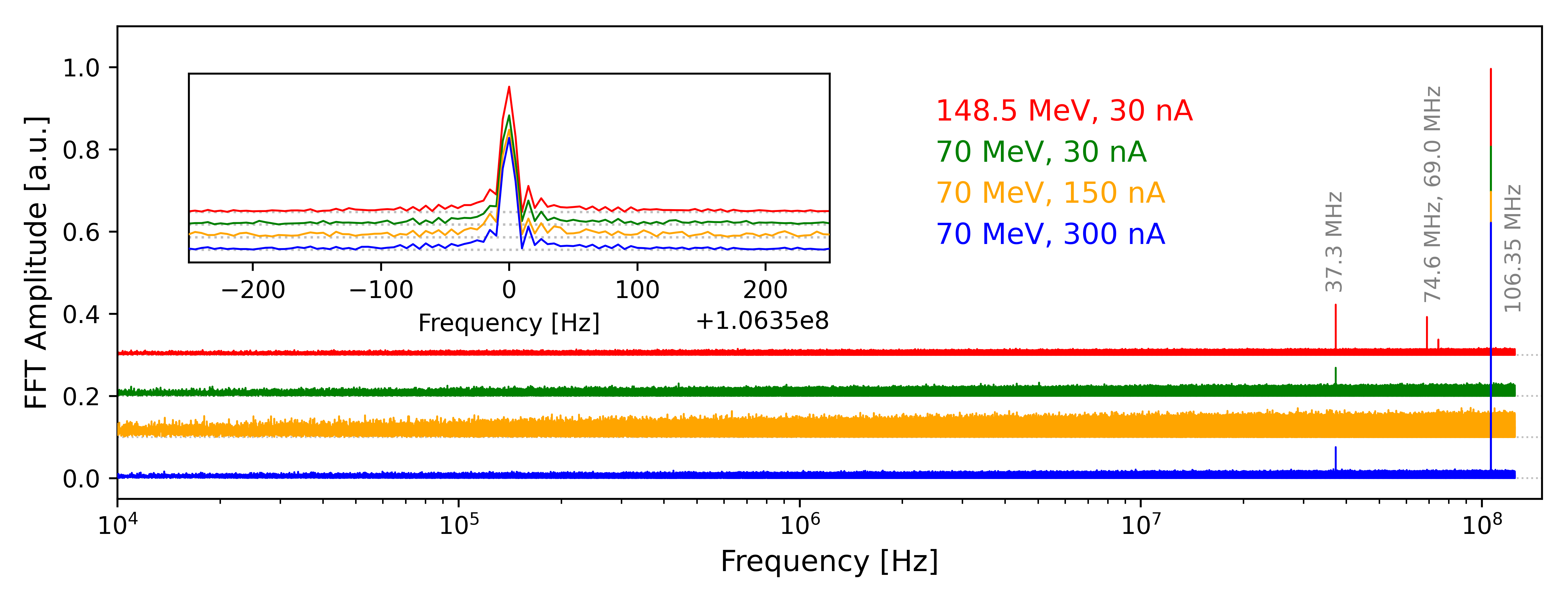}
\caption{Frequency spectra of the particle currents for the different proton beam settings at the Trento cyclotron. The spectra exhibit a single dominant resonance at the cyclotron \gls{RF} frequency (shown in the zoomed inset). \label{fig:FFT_Trento_all}}
\end{figure}

\begin{figure}[htbp]
\centering
\includegraphics[width=\textwidth]{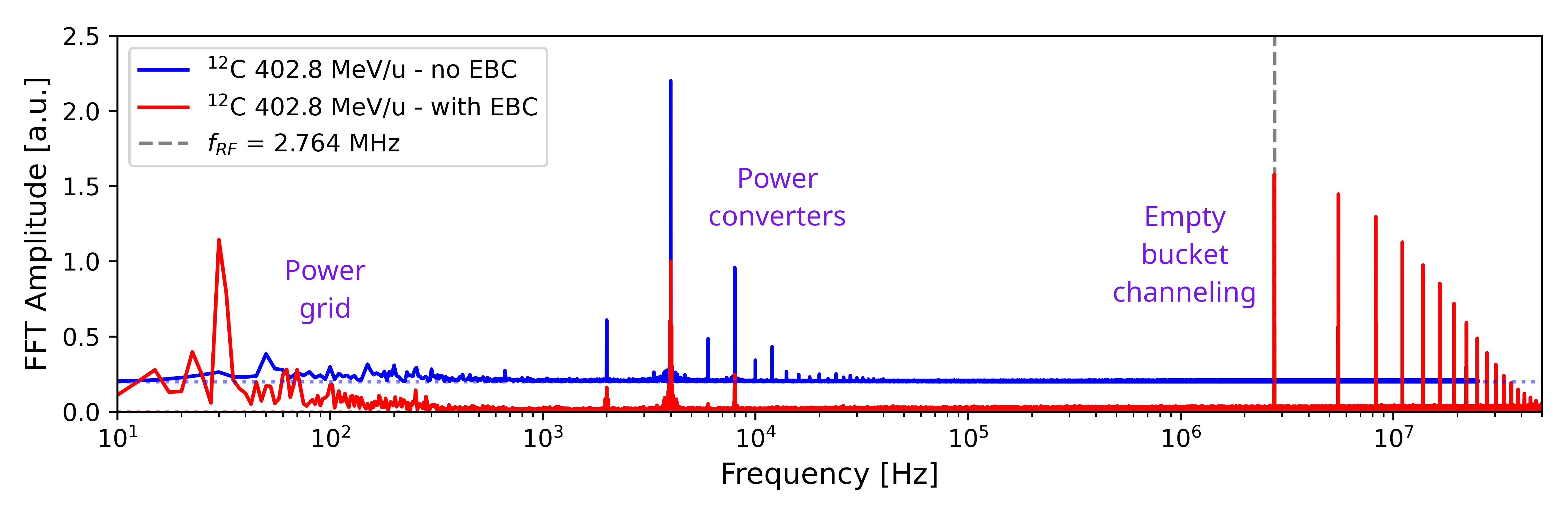}
\caption{Frequency spectra of the particle currents for \SI{402.8}{\mega\electronvolt\per\u} \ce{^12C^6+} beams at the MedAustron synchrotron with \gls{EBC} (blue) and without \gls{EBC} (red). The amplitudes for the measurement without \gls{EBC} have been scaled down for visibility. \label{fig:FFT_MedAustron_C402}}
\end{figure}

The registered particle current in the sensor is derived by integrating the detected event rate into specified time bins. Figure \ref{fig:MAUS_Intensity} shows the beam current for two full carbon spills at MedAustron with and without \gls{EBC}. Beam modulation can be characterized in the frequency domain via analysis of the registered particle currents. Figure \ref{fig:FFT_Trento_all} shows the frequency spectrum of the particle current for the measurements at the Trento cyclotron. Registered events per \SI{200}{\milli\second} acquisition have been integrated into \SI{4}{\nano\second} bins allowing for an analysis from \SI{5}{\hertz} to \SI{125}{\mega\hertz}. In all spectra, a single dominant peak is observed at \SI{106.35}{\mega\hertz}, shown in the zoomed inset in figure \ref{fig:FFT_Trento_all}, accompanied by minor resonances at \SI{37.3}{\mega\hertz} and \SI{74.6}{\mega\hertz}. Figure \ref{fig:FFT_MedAustron_C402} shows the frequency spectrum of the \SI{402.8}{\mega\electronvolt\per\u} \ce{^12C^6+} beam at MedAustron for the configuration with and without \gls{EBC}. The registered counts per \SI{200}{\milli\second} acquisitions have been integrated in \SI{10}{\nano\second} bins allowing for an analysis from \SI{5}{\hertz} to \SI{50}{\mega\hertz}. Due to the more complex accelerator design, the spectra exhibit several contributions that can be broadly categorized in three scales: a component around \SIrange[]{50}{100}{\hertz} associated with the power grid, ripples in the \SI{}{\kilo\hertz} regime from the magnet power converters with a prominent resonance at \SI{4}{\kilo\hertz} from the main ring dipole power converter, and, in the configuration with \gls{EBC}, a strong modulation at the \gls{RF} frequency and its harmonics. In figure \ref{fig:FFT_MedAustron_C402}, the amplitudes have been scaled for improved visibility since the relative strength of the modulation in the measurements without \gls{EBC} is substantially higher.

\section{Discussion}
\label{sec:Discussion}

The results for the \glspl{PATD} presented in section \ref{sec:Results}, which were measured using a known geometry can now be taken as a reference and scaled by detector surface area and the beamspot coverage to provide input and timing constraints for other experiments using devices in the beam. For simplicity, the \glspl{PATD} are fitted with an exponential envelope $\exp(\alpha \Delta t)$, enabling a rescaling of the distributions via $\alpha \rightarrow \alpha^\prime$. The intensity distribution of particles over the beamspot is modeled as an anisotropic 2D-Gaussian

\begin{equation}
    G(x,y) = \frac{1}{2 \pi \sigma_x \sigma_y} \exp\left( - \left[ \frac{x^2}{2\sigma_x^2} + \frac{y^2}{2\sigma_y^2} \right] \right).
\end{equation}

\begin{figure}[htbp]
\centering
\includegraphics[width=.47\textwidth]{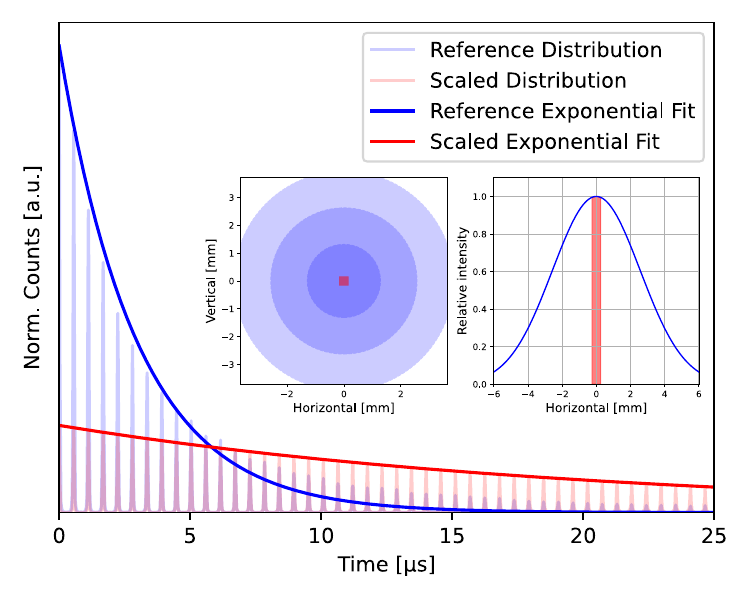}
\qquad
\includegraphics[width=.47\textwidth]{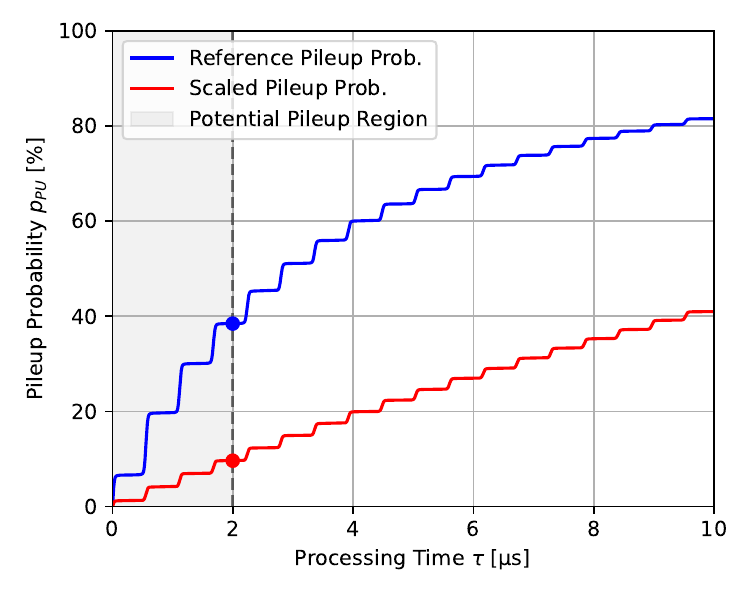}
\caption{The reference \glspl{PATD} and pileup probabilities (blue curves) can be rescaled to account for different detector geometries (red curves). The scaling factor is determined by the relative beamspot coverage of the detector areas, determined from \eqref{eq:integral}. The integration is illustrated in the left panel (top view and lateral profile of the beamspot). Combined with the processing time $\tau$ of a given readout system (e.g. \SI{2}{\micro\second}, indicated by the shaded region in the right plot), this gives an estimate of the expected pileup for that system, corresponding to inter-arrival times $\Delta t<\tau$.\label{fig:Pileup_scaling}}
\end{figure}

Assuming a square detector with side-lengths $L_x, L_y$, the envelopes $\alpha$ for the reference distribution and $\alpha^\prime$ for the rescaled distribution are proportional to the total average particle rate (over the entire beamspot) $R_\text{tot}$ weighted by the ratio $I_\text{Det} / I_\text{Spot}$ of the integrated intensity over the detector and over the full beamspot,

\begin{equation}
\label{eq:integral}
\begin{aligned}
    \alpha, \alpha^\prime \propto s, s^\prime &= R_\text{tot} \cdot \frac{I_\text{Det}}{I_\text{Spot}}
    = R_\text{tot}\cdot \frac{\int_{-L_x/2}^{L_x/2}\int_{-L_y/2}^{L_y/2} G(x,y)\, \mathrm{d}x\, \mathrm{d}y}
         {\int_{-\infty}^\infty \int_{-\infty}^\infty G(x,y)\, \mathrm{d}x\, \mathrm{d}y} = \\ &= R_\text{tot}\cdot \text{erf}\!\left( \frac{L_x}{2 \sqrt{2}\,\sigma_x} \right)
    \text{erf}\!\left( \frac{L_y}{2 \sqrt{2}\,\sigma_y} \right),
\end{aligned}
\end{equation}

with $\text{erf}(x)$ being the Gaussian error function. The scaling factor for the \glspl{PATD} is then calculated as the ratio $S=s^\prime/s=\alpha^\prime/\alpha$, with the absolute intensity $R_\text{tot}$ canceling out. For circular detectors, the calculation is analogous in polar coordinates. For more complex geometries, numerical integration is used. From the scaled \gls{PATD}, the pileup probability can be obtained via the cumulative sum. The beamspot geometry can typically be obtained from existing monitoring systems, archival data, or EBT3 film measurements. Figure \ref{fig:Pileup_scaling} illustrates the process.\\

Spectroscopic and timing sensitive measurement are often highly sensitive to pileup, such as in microdosimetry \cite{Knopf_2025_2} and ion imaging ~\cite{Ulrich_Pur_2024}. The operation of such systems in clinical facilities under high dose-rate conditions therefore requires dedicated electronics that account for the spill and micro-spill structure of the beams. The presented method provides a basis for pileup analysis using semiconductor detectors in such environments.

\section{Conclusion}
The micro-spill structure of accelerator beams typically deviates from a Poisson distributed time structure, as expected from a radioactive source, which needs to be considered in pileup estimations for pileup-sensitive measurements. The experimental determination of the micro-spill structure of the extracted beams using a fast readout system provides typically inaccessible information, relevant for researchers using medical cyclotrons and synchrotrons for physics experiments. It establishes quantitative constraints on the applicable timescales required for reliable event separation, thereby providing direct design guidance for new readout electronics. These results offer a basis for optimizing future hardware and experimental design with respect to timing performance and expected pileup conditions.

\acknowledgments
The authors want to thank Elisabeth Renner and Matthias Kausel from TU Wien and MedAustron for their help in the preparation of custom synchrotron settings and data extraction. This project has received funding from the Austrian Research Promotion Agency FFG, Austria, grants number 918092 and 925664. The financial support of the Austrian Ministry of Education, Science and Research is gratefully acknowledged for providing beam time and research infrastructure at MedAustron. The authors acknowledge TU Wien Bibliothek for financial support through its Open Access Funding Programme.

% Bibliography
\bibliographystyle{JHEP}
\bibliography{biblio.bib}

@article{Giordanengo_2013,
	title = {Design and characterization of the beam monitor detectors of the Italian National Center of Oncological Hadron-therapy ({CNAO})},
	volume = {698},
	issn = {0168-9002},
	doi = {10.1016/j.nima.2012.10.004},
	pages = {202--207},
	journal = {Nuclear Instruments and Methods in Physics Research Section A: Accelerators, Spectrometers, Detectors and Associated Equipment},
	shortjournal = {Nuclear Instruments and Methods in Physics Research Section A: Accelerators, Spectrometers, Detectors and Associated Equipment},
	author = {Giordanengo, S. and Donetti, M. and Garella, M.A. and Marchetto, F. and Alampi, G. and Ansarinejad, A. and Monaco, V. and Mucchi, M. and Pecka, I.A. and Peroni, C. and Sacchi, R. and Scalise, M. and Tomba, C. and Cirio, R.},
    year = {2013}
}

@article{Giordanengo_2015,
	title = {The {CNAO} dose delivery system for modulated scanning ion beam radiotherapy},
	volume = {42},
	issn = {2473-4209},
	doi = {10.1118/1.4903276},
	pages = {263--275},
	number = {1},
	journal = {Medical Physics},
	author = {Giordanengo, S. and Garella, M. A. and Marchetto, F. and Bourhaleb, F. and Ciocca, M. and Mirandola, A. and Monaco, V. and Hosseini, M. A. and Peroni, C. and Sacchi, R. and Cirio, R. and Donetti, M.},
	year = {2015}
}

@article{Ulrich-Pur_2021,
	title = {Commissioning of low particle flux for proton beams at {MedAustron}},
	volume = {1010},
	issn = {0168-9002},
	doi = {10.1016/j.nima.2021.165570},
	pages = {165570},
	journal = {Nuclear Instruments and Methods in Physics Research Section A: Accelerators, Spectrometers, Detectors and Associated Equipment},
	short = {Nuclear Instruments and Methods in Physics Research Section A: Accelerators, Spectrometers, Detectors and Associated Equipment},
	author = {Ulrich-Pur, Felix and Adler, Laurids and Bergauer, Thomas and Burker, Alexander and De Franco, Andrea and Guidoboni, Greta and Hirtl, Albert and Irmler, Christian and Kaser, Stefanie and Nowak, Sebastian and Pitters, Florian and Pivi, Mauro and Prokopovich, Dale and Schmitzer, Claus and Wastl, Alexander},
    year = {2021}
}

@article{Knopf_2025,
	title = {Characterizing the delivered spill structure of medical proton and carbon-ion beams at {MedAustron} using a high frequency silicon carbide readout},
    volume = {1082},
    issn = {0168-9002},
	doi = {10.1016/j.nima.2025.170984},
	pages = {170984},
	journal = {Nuclear Instruments and Methods in Physics Research Section A: Accelerators, Spectrometers, Detectors and Associated Equipment},
	short = {Nuclear Instruments and Methods in Physics Research Section A: Accelerators, Spectrometers, Detectors and Associated Equipment},
	author = {Knopf, Matthias and Gsponer, Andreas and Kausel, Matthias and Waid, Simon and Onder, Sebastian and Gundacker, Stefan and Radmanovac, Daniel and Magrin, Giulio and Bergauer, Thomas and Hirtl, Albert},
	year = {2025}
}

@article{DeFranco_2018,
	title = {Slow Extraction Optimization at the {MedAustron} Ion Therapy Center: Implementation of Front End Acceleration and {RF} Knock Out},
	volume = {{IPAC}2018},
	doi = {10.18429/JACOW-IPAC2018-MOPML025},
	journaltitle = {Proceedings of the 9th Int. Particle Accelerator Conf.},
	publisher = {{JACoW} Publishing, Geneva, Switzerland},
	author = {De Franco, Andrea and Adler, Laurids and Farinon, Fabio and Gambino, Nadia and Guidoboni, Greta and Kowarik, Gregor and Kronberger, Matthias and Kurfürst, Christoph and Myalski, Szymon and Nowak, Sebastian and Penescu, Liviu and Pivi, Mauro and Schmitzer, Claus and Strašík, Ivan and Urschütz, Peter and Wastl, Alexander},
	year = {2018}
}

@article{DeFranco_2021,
	title = {Optimization of synchrotron based ion beam therapy facilities for treatment time reduction, options and the {MedAustron} development roadmap},
	volume = {81},
	doi = {10.1016/j.ejmp.2020.11.029},
	pages = {264--272},
	journaltitle = {Physica Medica},
	author = {De Franco, Andrea and Schmitzer, Claus and Gambino, Nadia and Glatzl, Thomas and Myalski, Szymon and Pivi, Mauro},
	year = {2021}
}

@article{Knopf_2025_2,
    title = {Exploring offline pileup correction to improve the accuracy of microdosimetric characterization in clinical ion beams},
    volume = {70},
    number = {13},
    doi = {10.1088/1361-6560/ade6bc},
    pages = {135008},
    journal = {Physics in Medicine \& Biology},
    author = {Knopf, Matthias and Barna, Sandra and Radmanovac, Daniel and Bergauer, Thomas and Hirtl, Albert and Magrin, Giulio},
    year = {2025}
}

@article{DeNapoli_2022,
	title = {{SiC} detectors: A review on the use of silicon carbide as radiation detection material},
	volume = {10},
	issn = {2296-424X},
	doi = {10.3389/fphy.2022.898833},
	journal = {Frontiers in Physics},
	short = {Front. Phys.},
	author = {De Napoli, Marzio},
    year = {2022}
}

@article{Gsponer_2025,
	title = {Extraction of Electron and Hole Drift Velocities in Thin 4H-SiC PIN Detectors Using High-Frequency Readout Electronics},
	volume = {25},
	number = {23},
	issn = {1424-8220},
	doi = {10.3390/s25237196},
	journal = {Sensors},
	author = {Gsponer, Andreas and Onder, Sebastian and Gundacker, Stefan and Burin, Jürgen and Knopf, Matthias and Radmanovac, Daniel and Waid, Simon and Bergauer, Thomas},
	year = {2025}
}

@article{ICRU_93,
	title = {{ICRU} Report 93, Prescribing, Recording, and Reporting Light Ion Beam Therapy},
	volume = {16},
	issn = {1473-6691, 1742-3422},
	pages = {1--222},
	number = {1},
	journal = {Journal of the {ICRU}},
	shortjournal = {{ICRU}},
	author = {Jäkel, O. and Bert, C. and Fossati, P. and Kamada, T. and Karger, C.P. and Matsufuji, N. and Scholu, M.},
	year = {2019}
}

@article{CNAO,
	title = {The National Centre for Oncological Hadrontherapy ({CNAO}): Status and perspectives},
	volume = {31},
	issn = {1120-1797},
	doi = {10.1016/j.ejmp.2015.03.001},
	pages = {333--351},
	number = {4},
	journaltitle = {Physica Medica: European Journal of Medical Physics},
	author = {Rossi, Sandro},
	year = {2015}
}

@article{Trento,
	title = {Proton beam characterization in the experimental room of the Trento Proton Therapy facility},
	volume = {869},
	issn = {0168-9002},
	doi = {10.1016/j.nima.2017.06.017},
	pages = {15--20},
	journal = {Nuclear Instruments and Methods in Physics Research Section A: Accelerators, Spectrometers, Detectors and Associated Equipment},
	short = {Nuclear Instruments and Methods in Physics Research Section A: Accelerators, Spectrometers, Detectors and Associated Equipment},
	author = {Tommasino, F. and Rovituso, M. and Fabiano, S. and Piffer, S. and Manea, C. and Lorentini, S. and Lanzone, S. and Wang, Z. and Pasini, M. and Burger, W. J. and La Tessa, C. and Scifoni, E. and Schwarz, M. and Durante, M.},
	year = {2017}
}

@article{MedAustron,
    title = {MedAustron: First Years of Operation},
    volume = {29},
    number = {2},
    doi = {10.1080/10619127.2019.1603558},
    pages = {22--24},
    journal = {Nuclear Physics News},
    author = {T. Schreiner and M. Stock and P. Georg and D. Georg},
    year = {2019}
}

@misc{PTCOG,
  author       = {PTCOG},
  title        = {Particle Therapy Co-Operative Group: Particle therapy facilities in clinical operation},
  howpublished = {\url{https://www.ptcog.site/index.php/facilities-in-operation-public/}},
  year         = {Accessed: 15 March 2026}
}

@article{Review_Cyclotrons,
	title = {Comparison of cyclotron and synchrotron in particle therapy},
	volume = {5},
	issn = {2740-4218},
	doi = {10.1051/vcm/2024008},
	pages = {7},
	journaltitle = {Visualized Cancer Medicine},
	shortjournal = {Vis Cancer Med},
	author = {Xiao, Mei and Liu, Bing and Peng, Jingyu and Li, Mindi and Xie, Shuqing},
	year = {2024}
}

@phdthesis{Kuehteubl_Diss,
  author       = {Kühteubl, Florian},
  title        = {Slow Extraction Optimisation for the {MedAustron} Synchrotron},
  school       = {Technische Universität Wien},
  year         = {2024},
  address      = {Vienna},
  type         = {PhD thesis}
}

@techreport{Crescenti_1998_EBC,
      author        = "Crescenti, M",
      title         = "{{RF} empty bucket channelling combined with a betatron core
                       to improve slow extraction in medical synchrotrons}",
      institution   = "CERN",
      reportNumber  = "CERN-PS-97-068-DI",
      address       = "Geneva",
      year          = "1998",
      doi           = "https://cds.cern.ch/record/346139",
}

@book{PIMMS,
      author = "Badano, L and Benedikt, Michael and Bryant, P J and Crescenti, M and Holy, P and Maier, A T and Pullia, M and Rossi, S and Knaus, P",
      collaboration = "CERN-TERAFoundation-MedAustronOncology-2000",
      title         = "{Proton-Ion Medical Machine Study (PIMMS), 1}",
      year          = "2000",
      doi           = "https://cds.cern.ch/record/385378",
	  publisher 	= "CERN"
}

@article{Pivi_2019_Status_Carbon_Commissioning,
	title = {Status of the Carbon Commissioning and Roadmap Projects of the {MedAustron} Ion Therapy Center Accelerator},
	volume = {{IPAC}2019},
	doi = {10.18429/JACOW-IPAC2019-THXXPLS1},
	journal = {Proceedings of the 10th Int. Particle Accelerator Conf.},
	author = {Pivi, Mauro and Adler, Laurids and De Franco, Andrea and Farinon, Fabio and Gambino, Nadia and Guidoboni, Greta and Kowarik, Gregor and Kronberger, Matthias and Kurfürst, Christoph and Lau, Hin Tung and Myalski, Szymon and Nowak, Sebastian and Penescu, Liviu and Schmitzer, Claus and Strašík, Ivan and Urschütz, Peter and Wastl, Alexander},
	editor = {Mark (Ed.), Boland and Hitoshi (Ed.), Tanaka and David (Ed.), Button and Rohan (Ed.), Dowd and {RW} (Ed.), Volker, Schaa and Eugene (Ed.), Tan},
	year = {2019}
}

@article{Gambino_2024_Helium,
	title = {Status of helium ion beams commissioning at {MedAustron} ion therapy center},
	volume = {{IPAC}2024},
	doi = {10.18429/JACoW-IPAC2024-TUPS06},
	journaltitle = {{JACoW}},
	author = {Gambino, Nadia and Guidoboni, Greta and Kausel, Matthias and Pivi, Mauro and Plassard, Fabien and Rizzoglio, Valeria and Strasik, Ivan and Fischl, Lorenz and Penescu, Liviu and Prokopovich, Dale},
	urldate = {2025-07-07},
	year = {2024}
}

@article{Pullia_2016_Betatron_Extraction,
	title = {Betatron Core Driven Slow Extraction at {CNAO} and {MedAustron}},
    journal = {Proceedings of IPAC2016},
	author = {Pullia, M G and Bressi, E and Falbo, L and Priano, C and Rossi, S and Viviani, C and Foundation, {CNAO}},
	year = {2016}
}

@article{Ulrich_Pur_2024,
doi = {10.1088/1361-6560/ad3326},
year = {2024},
publisher = {IOP Publishing},
volume = {69},
number = {7},
pages = {075031},
author = {Ulrich-Pur, Felix and Bergauer, Thomas and Galatyuk, Tetyana and Hirtl, Albert and Kausel, Matthias and Kedych, Vadym and Kis, Mladen and Kozymka, Yevhen and Krüger, Wilhelm and Linev, Sergey and Michel, Jan and Pietraszko, Jerzy and Rost, Adrian and Schmidt, Christian Joachim and Träger, Michael and Traxler, Michael},
title = {First experimental time-of-flight-based proton radiography using low gain avalanche diodes},
journal = {Physics in Medicine \& Biology}
}

\end{document}